# Evaluation of Huffman and Arithmetic Algorithms for Multimedia Compression Standards

Asadollah Shahbahrami, Ramin Bahrampour, Mobin Sabbaghi Rostami,
Mostafa Ayoubi Mobarhan,
Department of Computer Engineering, Faculty of
Engineering, University of Guilan, Rasht, Iran.

E-mail: shahbahrami@guilan.ac.ir
ramin.fknr@gmail.com, mobin.sabbaghi@gmail.com,
 mostafa_finiks7@yahoo.com

**Abstract**

Compression is a technique to reduce the quantity of data without excessively reducing the quality of the multimedia data. The transition and storing of compressed multimedia data is much faster and more efficient than original uncompressed multimedia data. There are various techniques and standards for multimedia data compression, especially for image compression such as the JPEG and JPEG2000 standards. These standards consist of different functions such as color space conversion and entropy coding. Arithmetic and Huffman coding are normally used in the entropy coding phase. In this paper we try to answer the following question. Which entropy coding, arithmetic or Huffman, is more suitable compared to other from the compression ratio, performance, and implementation points of view? We have implemented and tested Huffman and arithmetic algorithms. Our implemented results show that compression ratio of arithmetic coding is better than Huffman coding, while the performance of the Huffman coding is higher than Arithmetic coding. In addition, implementation of Huffman coding is much easier than the Arithmetic coding.

## 1   Introduction

Multimedia data, especially images have been increasing every day. Because of their large capacity, storing and transmitting are not easy and they need large storage devices and high bandwidth network systems. In order to alleviate these requirements, compression techniques and standards such as JPEG, JPEG2000, MPEG-2, and MPEG-4 have been used and proposed. To compress something means that you have a piece of data and you decrease its size [1, 2, 3, 4, 5]. The JPEG is a well-known standardized image compression technique that it loses information, so the decompressed picture is not the same as the original one. Of course the degree of losses can be adjusted by setting the compression parameters. The JPEG standard constructed from several functions such as DCT, quantization, and entropy coding. Huffman and arithmetic coding are the two most important entropy coding in image compression standards. In this paper, we are planning to answer the following question. Which entropy coding, arithmetic or Huffman, is more suitable from the compression ratio, performance, and implementation points of view compared to other?

 We have studied, implemented, and tested these important algorithms using different image contents and sizes. Our experimental results show that compression ratio of arithmetic coding is higher than Huffman coding, while the performance of the Huffman coding is higher than Arithmetic coding. In addition, implementation complexity of Huffman coding is less than the Arithmetic coding.

The rest of the paper is organized as follow. Section 2 describes the JPEG compression standard and Section 3 and 4 explain Huffman and arithmetic algorithms, respectively. Section 5 discusses implementation of the algorithms and standard test images. Experimental results are explained in Section 6 followed by related work in Section 7. Finally, conclusions are drawn in Section 8.

## 2    The JPEG Compression Standard

The JPEG is an image compression standard developed by the Joint Photographic Experts Group. It was formally accepted as an international in 1992. The JPEG consists of a number of steps, each of which contributes to compression [3].

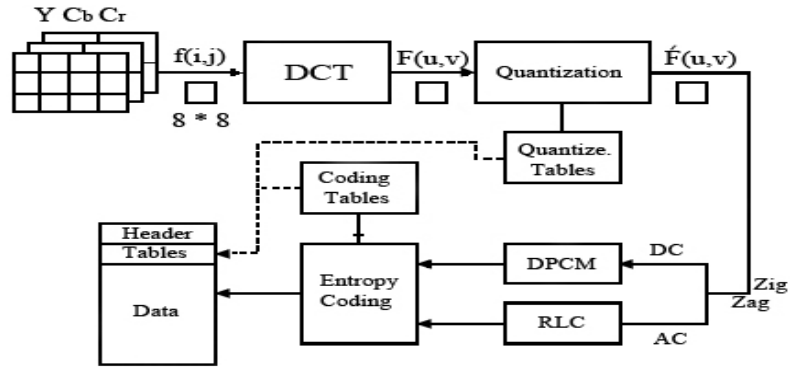

Figure 1. Block diagram of the JPEG encoder [3].

Figure 1 shows a block diagram for a JPEG encoder. If we reverse the arrows in the figure, we basically obtain a JPEG decoder. The JPEG encoder consists of the following main steps.

The first step is about color space conversion. Many color images are represented using the RGB color space. RGB representations, however, are highly correlated, which implies that the RGB color space is not well-suited for independent coding [29]. Since the human visual system is less sensitive to the position and motion of color than luminance [6, 7]. Therefore, some color space conversions such as RGB to YCbCr are used [29, 8]. The next step of the JPEG standard consists of Discrete Cosine Transform (DCT). A DCT expresses a sequence of finitely many data points in terms of a sum of cosine functions oscillating at different frequencies. DCTs are an important part in numerous applications in science and engineering for the lossless compression of multimedia data [1, 3]. The DCT separates the image into different frequencies part. Higher frequencies represent quick changes between image pixels and low frequencies represent gradual changes between image pixels. In order to perform the DCT on an image, the image should be divided into $8 \times 8$ or $16 \times 16$ blocks [9].

In order to keep some important DCT coefficients, quantization is applied on the transformed block [10, 11]. After this step zigzag scanning is used. There are many runs of zeros in an image which has been quantized throughout the matrix so, the $8 \times 8$ blocks are reordered as single 64-element columns [4, 9]. We get a vector sorted by the criteria of the spatial frequency that gives long runs of zeros. The DC coefficient is treated separately from the 63 AC coefficients. The DC coefficient is a measure of the average value of the 64 image samples [12].

Finally, in the final phases coding algorithms such as Run Length Coding (RLC) and Differential Pulse Code Modulation (DPCM) and entropy coding are applied. The RLC is a simple and popular data compression algorithm [13]. It is based on the idea to replace a long sequence of the same symbol by a shorter sequence. The DC coefficients are coded separately from the AC ones. A DC coefficient is coded by the DPCM, which is a lossless data compression technique. While AC coefficients are coded using RLC algorithm. The DPCM algorithm records the difference between the DC coefficients of the current block and the previous block [14]. Since there is usually strong correlation between the DC coefficients of adjacent 8×8 blocks, it results a set of similar numbers with high occurrence [15]. DPCM conducted on pixels with correlation between successive samples leads to good compression ratios [16]. Entropy coding achieves additional compression using encoding the quantized DCT coefficients more compactly based on their statistical characteristics. Basically entropy coding is a critical step of the JPEG standard as all past steps depend on entropy coding and it is important which algorithm is used, [17]. The JPEG proposal specifies two entropy coding algorithms, Huffman [18] and arithmetic coding [19]. In order to determine which entropy coding is suitable from performance, compression ratio, and implementation points of view, we focus on the mentioned algorithms in this paper.

## 3   Huffman Coding

In computer science and information theory, Huffman coding is an entropy encoding algorithm used for lossless data compression [9]. The term refers to the use of a variable-length code table for encoding a source symbol (such as a character in a file) where the variable-length code table has been derived in a particular way based on the estimated probability of occurrence for each possible value of the source symbol. Huffman coding is based on frequency of occurrence of a data item. The principle is to use a lower number of bits to encode the data that occurs more frequently [1]. The average length of a Huffman code depends on the statistical frequency with which the source produces each symbol from its alphabet. A Huffman code dictionary [3], which associates each data symbol with a codeword, has the property that no code-word in the dictionary is a prefix of any other codeword in the dictionary [20]. The basis for this coding is a code tree according to Huffman, which assigns short code words to symbols frequently used and long code words to symbols rarely used for both DC and AC coefficients, each symbol is encoded with a variable-length code from the Huffman table set assigned to the 8x8 block's image component. Huffman codes must be specified externally as an input to JPEG encoders. Note that the form in which Huffman tables are represented in the data stream is an indirect specification with which the decoder must construct the tables themselves prior to decompression [4]. The algorithm for building the encoding follows this algorithm each symbol is a leaf and a root.  The flowchart of the Huffman algorithm is depicted in figure 2.

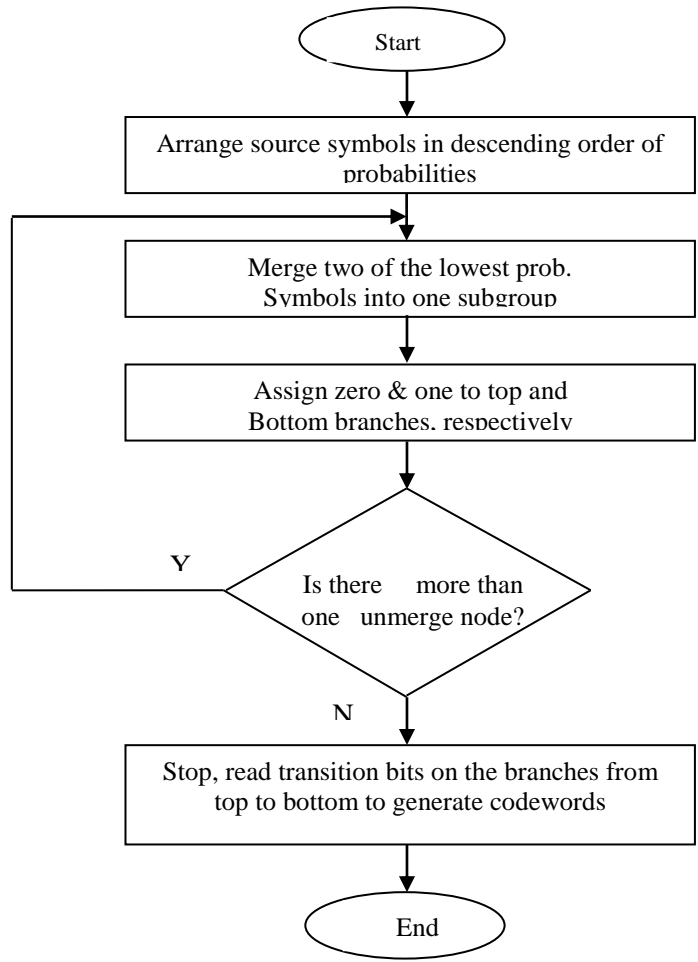

Figure 2. The flowchart of Huffman algorithm.

In order to clarify this algorithm, we give an example. We suppose that a list consists of 0, 2, 14, 136, and 222 symbols. Their occurrences are depicted in Table 1. As this table shows, symbol 0 occurs 100 times in the mentioned list. The Huffman tree and their final code are shown in figure 3 and Table 2 [21, 3].

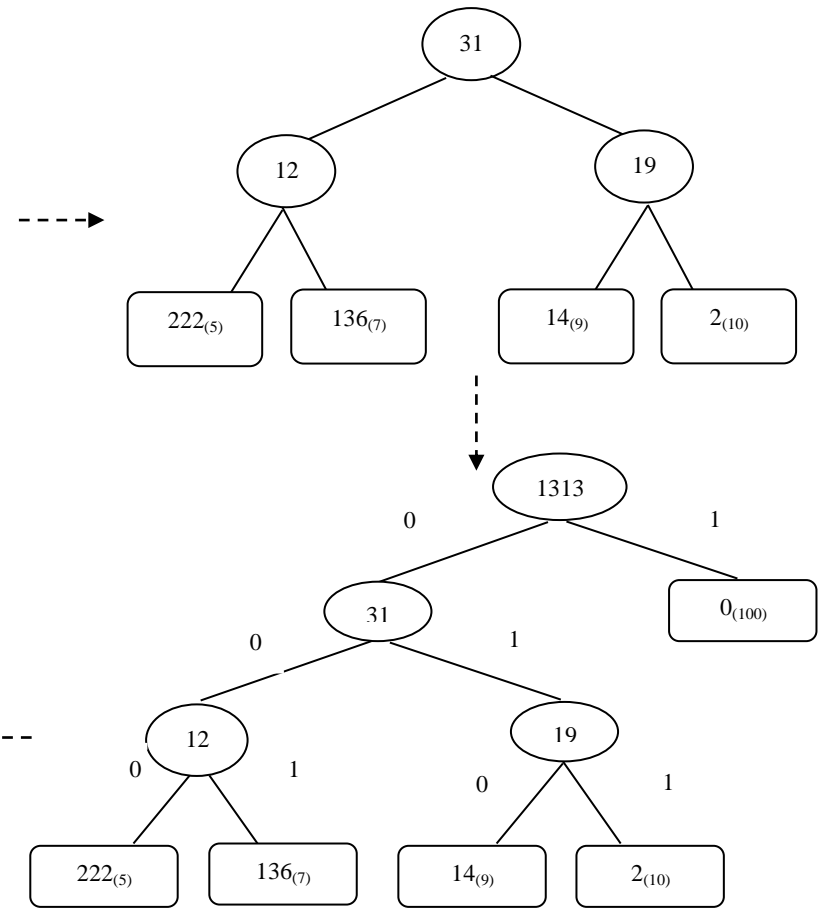

| Symbols | Frequency |
|---|---|
| 0 | 100 |
| 2 | 10 |
| 14 | 9 |
| 136 | 7 |
| 222 | 5 |

Table 1. Input symbols with their frequency.

| Symbols | Code | Frequency |
|---|---|---|
| 0 | 1 | 100 |
| 2 | 011 | 10 |
| 14 | 010 | 9 |
| 136 | 001 | 7 |
| 222 | 000 | 5 |

Table 2. Sequence of symbols and codes that are sent to the decoders.

Figure 3. Process of building Huffman tree.

As can be seen in Table 2, the minimum number of bits that is assigned to the largest occurrences symbol is one bit, bit 1 that is assigned to symbol 0. This means that we cannot assign fewer bits than one bit to that symbol. This is the main limitation of the of the Huffman coding. In order to overcome on this problem arithmetic coding is used that is discussed in the following section.

## 4   Arithmetic Coding

Arithmetic coding assigns a sequence of bits to a message, a sting of symbols. Arithmetic coding can treat the whole symbols in a list or in a message as one unit [22]. Unlike Huffman coding, arithmetic coding doesn´t use a discrete number of bits for each. The number of bits used to encode each symbol varies according to the probability assigned to that symbol. Low probability symbols use many bit, high probability symbols use fewer bits [23]. The main idea behind Arithmetic coding is to assign each symbol an interval. Starting with the interval [0...1), each interval is divided in several subinterval, which its sizes are proportional to the current probability of the corresponding symbols [24]. The subinterval from the coded symbol is then taken as the interval for the next symbol. The output is the interval of the last symbol [1, 3]. Arithmetic coding algorithm is shown in the following.

```
BEGIN
 low = 0.0; high = 1.0; range = 1.0;
 while (symbol != terminator)
  {   get (symbol);
       low = low + range * Range_low(symbol);
        high = low + range * Range_high(symbol);
         range = high - low;     }
    output a code so that low <= code < high;
END.[3]
```

The Figure 4 depicts the flowchart of the arithmetic coding.

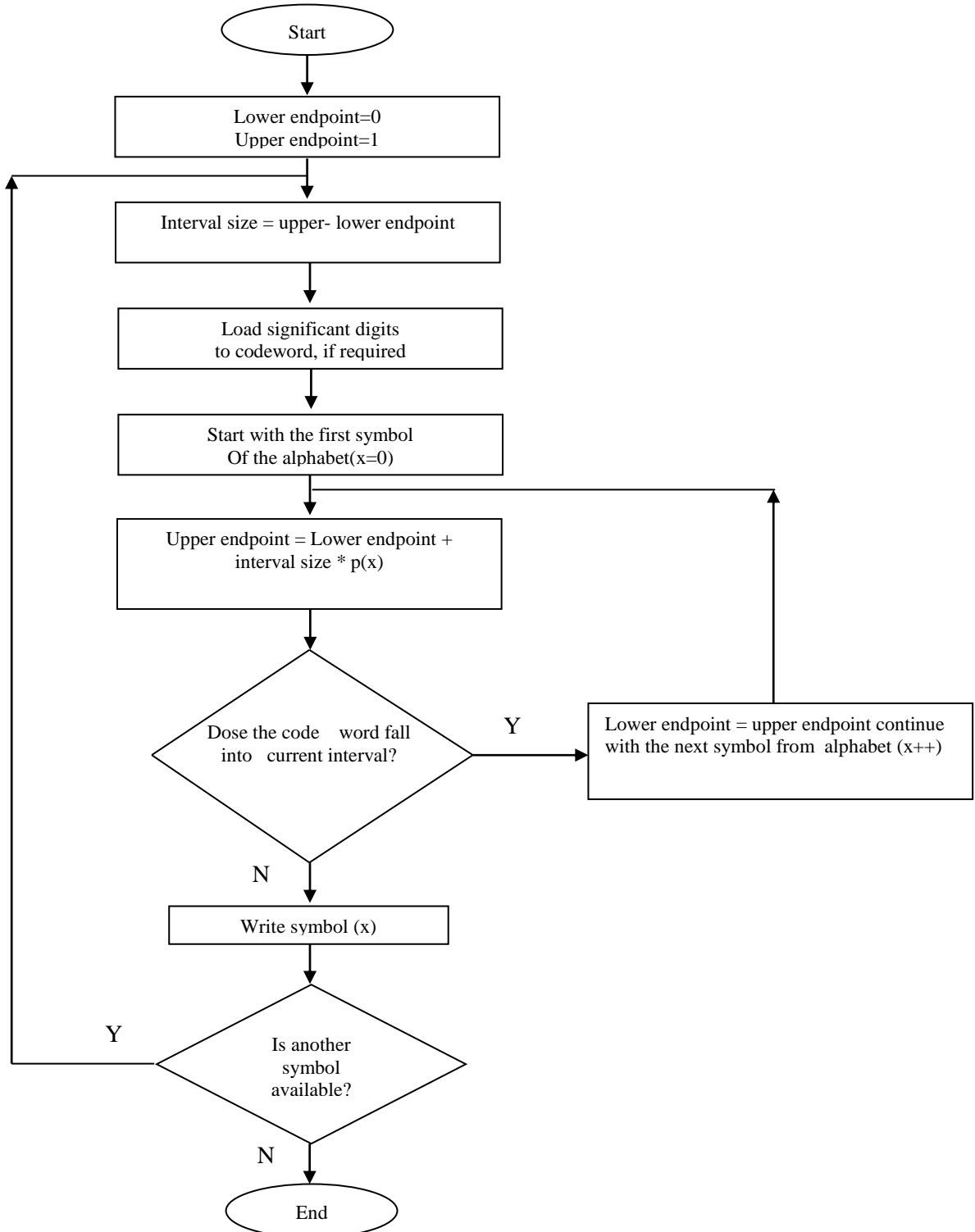

Figure 4. The flowchart of the arithmetic algorithm.

In order to clarify the arithmetic coding, we explain the previous example using this algorithm. Table 3 depicts the probability and the range of the probability of the symbols between 0 and 1.

| Symbols | Probability | Range |
|---|---|---|
| 0 | 0.63 | [ 0 , 0.63 ) |
| 2 | 0.11 | [ 0.63 , 0.74 ) |
| 14 | 0.1 | [ 0.74 , 0.84 ) |
| 136 | 0.1 | [ 0.84 , 0.94 ) |
| 222 | 0.06 | [ 0.94 , 1.0 ) |

Table 3. Probability and ranges distribution of symbols

We suppose that the input message consists of the following symbols: 2 0 0 136 0 and it start from left to right. Figure 5 depicts the graphical explanation of the arithmetic algorithm of this message from left to right. As can be seen, the first probability range is 0.63 to 0.74 (Table 3) because the first symbol is 2.

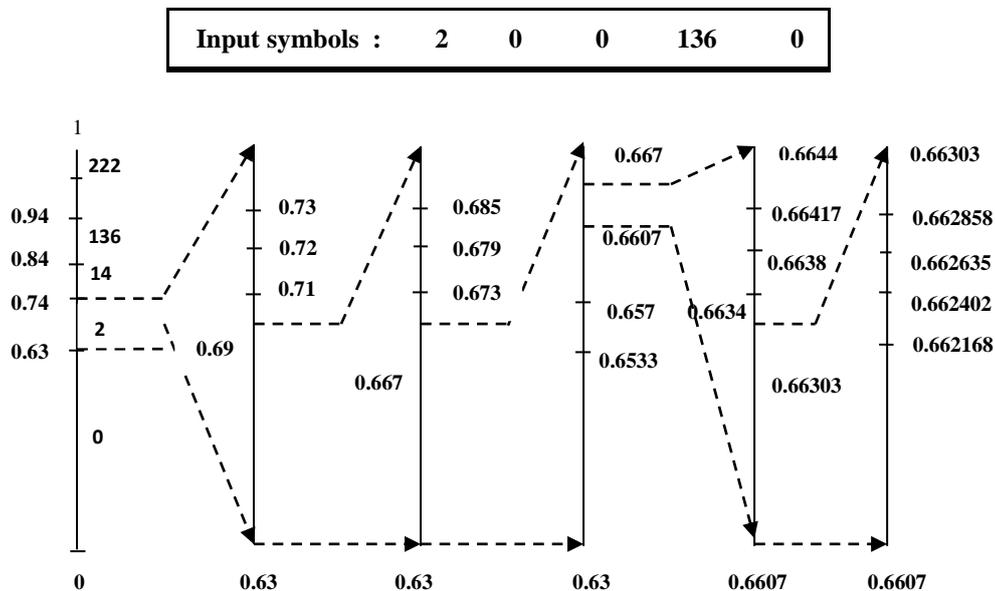

Figure 5. Graphical display of shrinking ranges.

Output : [0.6607 , 0.66303 )

The encoded interval for the mentioned example is [0.6607, 0.66303). A sequence of bits are assigned to a number that is located in this range.

Referring to Figure 2 and 4 and considering the discussed example in Figure 3 and 5, we can say that implementation complexity of arithmetic coding is more than Huffman. We saw this behavior in the programming too.

## 5  Implementation of the Algorithms

We have implemented Huffman and arithmetic algorithms using Matlab programming tools. We executed the implemented programs in a platform that its specification is depicted in Table 4.

| Laptop DELL XPS 1558 | |
|---|---|
| **RAM** | DDR3 - 4GB |
| **Processor Type** | i7-820QM |
| **Number of Cores of Processor** | 4 |
| **Clock Speed of Processor** | 1.73 GHz |
| **Cache of Processor** | 8 MB |
| **Operating System** | Windows 7 - 64bit |

Table 4: Specification of the platform system that has been used for execution of the programs.

A part of the implemented codes is depicted in Figure 6. We executed and tested both codes on many standard and famous images such as "Lena image". These standard test images have been used by different researchers [25, 26, 27, 28] related to image compression and image applications. We use different image sizes such as 128×128, 256×256, 512×512, 1024×1024 and 2048×2048. The same inputs are used for both algorithms.

```
%***********Start Huffman Coding
for time= 1:100
tic
k=0;
VECTOR-HUFF(1) = V(1);
  for l= 1:m
       a=0;
       for q=1:k
       if(VECTOR (l == VECTOR-HUFF (q));
            a=a+1;
       end
       end
       if (a==0)
            k=k+1;
            VECTOR-HUFF(k) = V(l);
       end
  end
  for u=1:k
       a=0;
       for l=1:m
       if (V(l)== VECTOR-ARITH(u))
            a=a+1;
       end
            VECTOR-HUFF-NUM(u)= a;
       end
  end
  for i=1:k
  P(i)= VECTOR-HUFF-NUM (i)/(m1);
  end
dict = huffmandict(VECTOR-HUFF,P);
hcode = huffmanenco(VECTOR,dict);
[f1,f2] = size(hcode);
Compression ratio = b0/f2
toc
end
```

```
%*********Start Arithmetic Coding
for time= 1:100
tic
k=0;
VECTOR-ARITH(1) = V(1);
  for l= 1:m
       a=0;
       for q=1:k
       if (V(l) == VECTOR-ARITH (q));
            a=a+1;
       end
       end
       if (a==0)
            k=k+1;
            VECTOR-ARITH (k) = V(l);
       end
  end
  for u=1:k
       a=0;
       for l=1:m
       if (V (l)== VECTOR-ARITH (u))
            a=a+1;
            Varith(l)=u;
       end
            VECTOR-ARITH-NUM(u)= a;
       end
  end
code = arithenco(Varith,VECTOR-ARITH-NUM);
[f1,f2] = size(code);
Compression ratio = b0/f2
toc
end
```

Figure 6. The segment codes of entropy coding.

# 6 Experimental Results

The experimental results of the implemented algorithms, Huffman and arithmetic coding for compression ratio and execution time are depicted in Table 5. As this table shows, on one hand, the compression ratio of the arithmetic coding for different image sizes is higher than the Huffman coding. On the other hand, arithmetic coding needs more execution time than Huffman coding. This means that the high compression ratio of the arithmetic algorithm is not free. It needs more resources than Huffman algorithm.

| Test Image Size | Compression Ratio (bits/sample) | | Algorithm Execution Times(seconds) | | Comparison Arithmetic to Huffman (%) | |
| --- | --- | --- | --- | --- | --- | --- |
| | Huffman | Arithmetic | Huffman | Arithmetic | Compression | Time |
| *2048 × 2048* | 6.37 | 12.02 | 32.67 | 63.22 | 47 | 48 |
| *1024 × 1024* | 5.64 | 7.73 | 8.42 | 20.37 | 27 | 58 |
| *512 × 512* | 5.27 | 6.55 | 2.13 | 5.67 | 19 | 59 |
| *256 × 256* | 4.78 | 5.40 | 0.55 | 1.63 | 11 | 66 |
| *128 × 128* | 4.38 | 4.65 | 0.14 | 0.45 | 5 | 68 |

Table 5. Average of compression results on test image set.

Another behavior that can be seen in Table 5 is, by increasing image sizes from 128X128 to 2048X2048, the improvement of the compression ratio of the arithmetic coding increases more than the Huffman coding. For instance, the compression ratio of Huffman algorithm for image sizes of 1024X1024 and 2048X2048 is 5.64 and 6.37, respectively. While for arithmetic coding is 7.73 and 12.02, respectively. Figures 7 and 8 depict a comparison of the compression ratio and execution time for the arithmetic and Huffman algorithms, respectively. In other words, these figures are the other representation of presented results in Table 5.

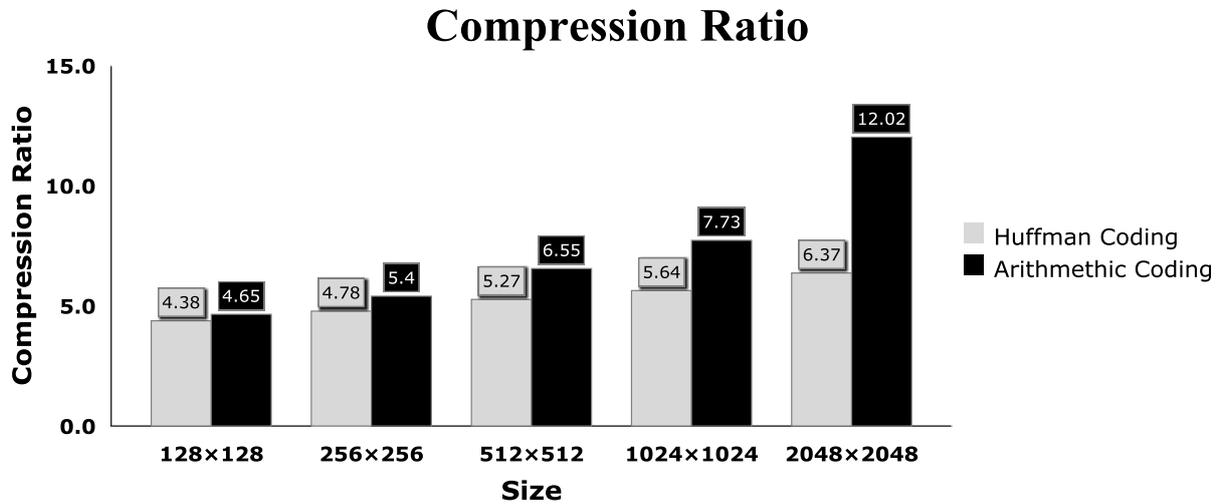

Figure 7. Comparison of compression ratio for Huffman and arithmetic algorithms using different image sizes.

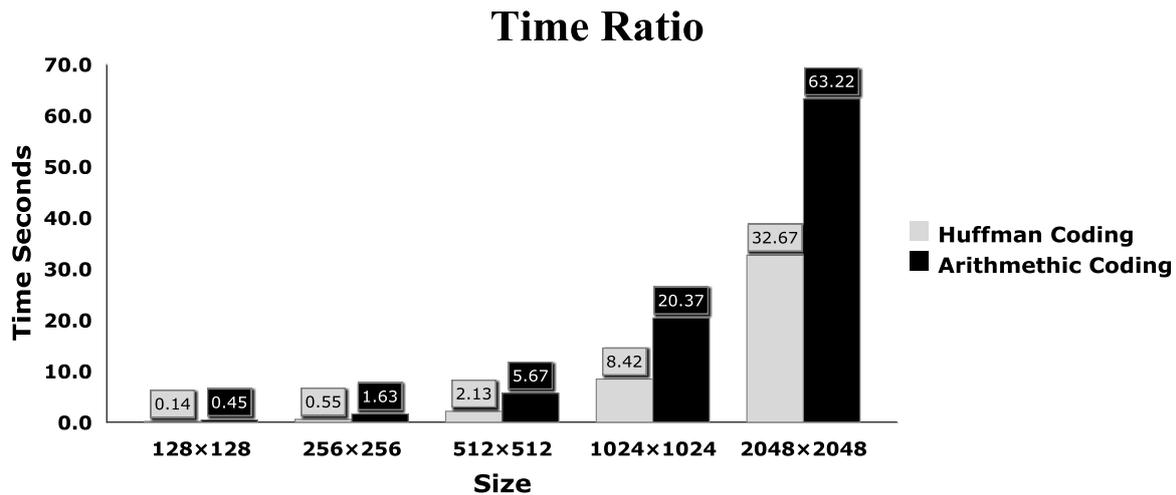

Figure 8. Comparison of performance for Huffman and arithmetic algorithms using different image sizes.

## 7- Related Work

Huffman[18] in 1952 proposed an elegant sequential algorithm which generates optimal prefix codes in $O(n\log n)$ time. The algorithm actually needs only linear time provided that the frequencies of appearances are sorted in advance. There have been extensive researches on analysis, implementation issues and improvements of the Huffman coding theory in a variety of applications [31, 32]. In [33], a two-phase parallel algorithm for time efficient construction of Huffman codes has been proposed. A new multimedia functional unit for general-purpose processors has been proposed in [34] in order to increase the performance of Huffamn coding.

Texts are always compressed with lossless compression algorithms. This is because a loss in a text will change its original concept. Repeated data is important in text compression. If a text has many repeated data, it can be compressed to a high ratio. This is due to the fact that compression algorithms generally eliminate repeated data. In order to evaluate the compression algorithms on the text data, a comparison between arithmetic and Huffman coding algorithms for different text files with different capacities has been performed in [30]. Experimental results showed that the compression ratio of the arithmetic coding for text files is better than Huffamn coding, while the performance of the Huffman coding is better than the arithmetic coding.

## 8- Conclusions

Compression is an important technique in the multimedia computing field. This is because we can reduce the size of data and transmitting and storing the reduced data on the Internet and storage devices are faster and cheaper than uncompressed data. Many image and video compression standards such as JPEG, JPEG2000, and MPEG-2, and MPEG-4 have been proposed and implemented. In all of them entropy coding, arithmetic and Huffman algorithms are almost used. In other words, these algorithms are important parts of the multimedia data compression standards. In this paper we have focused on these algorithms in order to clarify their differences from different points of view such as implementation, compression ratio, and performance. We have explained these algorithms in detail, implemented, and tested using different image sizes and contents. From implementation point of view, Huffman coding is easier than arithmetic coding. Arithmetic algorithm yields much more compression ratio than Huffman algorithm while Huffman coding needs less execution time than the arithmetic coding. This means that in some applications that time is not so important we can use arithmetic algorithm to achieve high compression ratio, while for some applications that time is important such as real-time applications, Huffman algorithm can be used.

In order to achieve much more performance compared to software implementation, both algorithms can be implemented on hardware platform such as FPGAs using parallel processing techniques. This is our future work.